\documentclass[a4paper,11pt]{article}
\pdfoutput=1 

\usepackage{jheppub} 
\usepackage{color}
\usepackage[T1]{fontenc} 
\usepackage{graphicx}
\usepackage{epsfig}

\title{\boldmath Locate QCD Critical End Point in a Continuum Model Study}

\author[1,2]{Chao Shi}
\author[3,4]{Yong-long Wang}
\author[5,2]{Yu Jiang}
\author[3,2]{Zhu-fang Cui}
\author[3,2,6,a]{and Hong-Shi Zong\note{Corresponding author.}}

\affiliation[1]{Key Laboratory of Modern Acoustics, MOE, Institute of Acoustics, and Department of Physics, Nanjing University, Nanjing 210093, China}
\affiliation[2]{State Key Laboratory of Theoretical Physics, Institute of Theoretical Physics, CAS, Beijing 100190, China}
\affiliation[3]{Department of Physics, Nanjing University,Nanjing 210093, China}
\affiliation[4]{Department of Physics, School of Science, Linyi University, Linyi 276005, P. R. China}
\affiliation[5]{Center for Statistical and Theoretical Condensed Matter Physics, Zhejiang Normal University, Jinhua City, Zhejiang Province 321004, China}
\affiliation[6]{Joint Center for Particle, Nuclear Physics and Cosmology, Nanjing 210093, China}

\emailAdd{shichao@chenwang.nju.edu.cn}
\emailAdd{wylong322@163.com}
\emailAdd{jiangyu@zjnu.cn}
\emailAdd{njucui@163.com}
\emailAdd{zonghs@chenwang.nju.edu.cn}

\abstract{With a modified chemical potential dependent effective model for the gluon propagator, we try to locate the critical end point (CEP) of strongly interacting matter in the framework of Dyson-Schwinger equations (DSE). Beyond the chiral limit, we find that Nambu solution and Wigner solution could coexist in some area. Using the Cornwall-Jackiw-Tomboulis (CJT) effective action, we show that these two phases are connected by a first order phase transition. We then locate CEP as the end point of the first order phase transition line. Meanwhile, based on CJT effective action, we give a direct calculation for the chiral susceptibility and thereby study the crossover.}


\begin{document}
\maketitle
\flushbottom

\section{Introduction}
\label{sec:intro}
Quantum chromodynamics (QCD) has two important features, namely, dynamical chiral symmetry breaking (D$\chi$SB) and color confinement. It is generally believed that with increasing temperature or baryon number density strongly interacting matter will undergo a phase transition from the hadronic matter to the quark-gluon plasma (QGP) which is expected to appear in the ultrarelativistic heavy ion collisions. These two phases are generally referred to as the Nambu-Goldstone phase and the Wigner phase concerning its realization of chiral symmetry. A central goal of the worldwide program in relativistic heavy ion collisions is to chart the phase diagram of QCD in the plane of nonzero temperature ($T$) and chemical potential ($\mu$) and to locate the position of the critical end point (CEP) \cite{a,c,v}.

In the Nambu-Goldstone phase the quarks are confined and chiral symmetry is dynamically broken which leads to a large dynamical quark mass. In the Wigner phase which is thought to be connected with QGP, chiral symmetry is partially restored and quarks are not confined. Theoretically, these two phases are described by two different solutions, the Nambu-Goldstone solution and the Wigner solution of the dressed quark propagator. The existence and properties of these two solutions in the chiral limit (i.e., the current quark mass $m=0$) have been widely investigated in the framework of Dyson-Schwinger equations (DSE) approach of QCD. For example, in Ref. \cite{d} the authors studied this problem by taking the infrared part of the gluon propagator in the Maris-Tandy model \cite{e} and inputting it directly into DSE at finite temperature and chemical potential. It is found that in the chiral limit there is an area on the $T-\mu$ plane on which both the Nambu and Wigner solutions coexist. However, there remain two questions. Firstly, in the real world, the current quark masses ($m$) for the $u$ and $d$ quarks are small but nonzero. Actually, many studies have shown that the absence of current quark mass would exhibits a second order phase transition instead of a crossover in certain areas of QCD phase diagram, which results in giving a tri-critical end point(TEP) instead of a CEP \cite{w,g}. Here, to be closer to the real world, we would take $m=5$ MeV. Secondly, in earlier work employing the Maris-Tandy model one simply ignores the $\mu$-dependence of the gluon propagator. However, when $\mu$ is large, the influence of $\mu$ to the gluon propagator should be important. Recently, authors of Refs. \cite{h,f1,f2,j} have noticed this problem and modified the gluon propagator. They have added terms in their own way to suppress the influence of the gluon propagator at large $\mu$. It is reasonable since it characterizes the weakening of interaction between quarks as $\mu$ goes up. Using such a modified gluon propagator, the authors of Ref. \cite{i} have found multi-solution of the quark gap equation at $T=0$ and $\mu\neq 0$ beyond the chiral limit (Similarly, the authors of Refs. \cite{j12,c13,c14} studied the multi-solutions of the quark gap equation with nonzero current quark mass in the case of finite temperature and chemical potential by introducing some modification to the normal Nambu-Jona-Lasinio model). The main purpose of this paper is to generalize the study in Ref. \cite{i} to the case of nonzero $T$ and $\mu$ and thereby locate the position of CEP by means of DSE.

In this work, we only consider two degenerate light flavors $u$ and $d$, and hence $\mu=\mu_u=\mu_d\approx\frac{1}{3}\mu_B$ throughout this paper.

\section{Gap equation and its solutions}
\label{eqs}
In Euclidean space and under rainbow-ladder approximation,the quark gap equation at finite $T$ and $\mu$ reads
\begin{equation}
\label{eq:a}
G^{-1}(\vec{p},\tilde{\omega}_n)=G_0^{-1}(\vec{p},
\tilde{\omega}_n)+\frac{4}{3}T\sum_{l=-\infty}^\infty\int\frac{d^3q}{(2\pi)^3}g^2D_{\mu \nu}(\vec{p}\!-\!\vec{q},\tilde{\omega}_n\!\!-\!\tilde{\omega}_l)\gamma_{\mu}G(\vec{q},\tilde{\omega}_l)\gamma_{\nu},
\end{equation}
where $\tilde{\omega}_n$=$(2n+1)\pi T+i\mu$.

Here we have put the regularization mass scale at infinity so that all renormalization constants are 1. This is allowed since the model used by us renders all the integrations convergent. Here we note that $G^{-1}(\vec{p},\tilde{\omega}_n)$ also depends on $T$ and $\mu$. It can be expressed as \cite{i,l}:
\begin{equation}
\label{eq:b}
G^{-1}(\vec{p},\tilde{\omega}_n;T,\mu)=i\vec{\gamma}\cdot\vec{p}A(\vec{p}^{~\!2},\tilde{\omega}_n^2;T,\mu)+i\gamma_4\tilde{\omega}_nC(\vec{p}^{~\!2},\tilde{\omega}_n^2;T,\mu)+B(\vec{p}^{~\!2},\tilde{\omega}_n^2;T,\mu).
\end{equation}

Substituting Eq.~\eqref{eq:b} into  Eq.~\eqref{eq:a}, one obtains the coupled integral equations for complex scalar functions $A$, $B$ and $C$. Here we need to specify the form of the model gluon propagator $g^2D_{\mu\nu}(Q)$ with $Q=(\vec{p}\!-\!\vec{q},\tilde{\omega}_n\!\!-\!\tilde{\omega}_l)$. In this work we adopt the model gluon propagator proposed in \cite{h,i}
\begin{equation}
\label{eq:MT}
g^2D_{\mu\nu}(Q;T,\mu)=\frac{4\pi^2}{\omega^6}D_0e^{-(Q^2+\alpha\mu^2)/\omega^2}(Q^2\delta_{\mu\nu}-Q_\mu Q_\nu).
\end{equation}
This model is based on the Maris-Tandy model, which is successful in describing the properties of mesons when used with the rainbow-ladder approximation, especially for the ground state pseudo-scalar and vector mesons \cite{e,m}. The form of this model entails that we are working in the Landau gauge, which is a fixed point of the renormalisation group. It has been used widely by Dyson-Schwinger equation studies both in hadron physics and finite-temperature QCD  \cite{l}. The parameters $D_0$ and $\omega$ can be fixed by the masses and electro-weak decay constants of the pions and $\rho$ mesons. However, it was found that for $\omega$ between $[0.3,0.5]$ GeV, $D_0$ which satisfies $\omega D_0=(0.8 \textrm{GeV})^3$ gave almost the same results \cite{y}. Later in this paper, we will fix $\omega$ further.

\begin{figure}
\centering 
\includegraphics[width=\textwidth,origin=c] {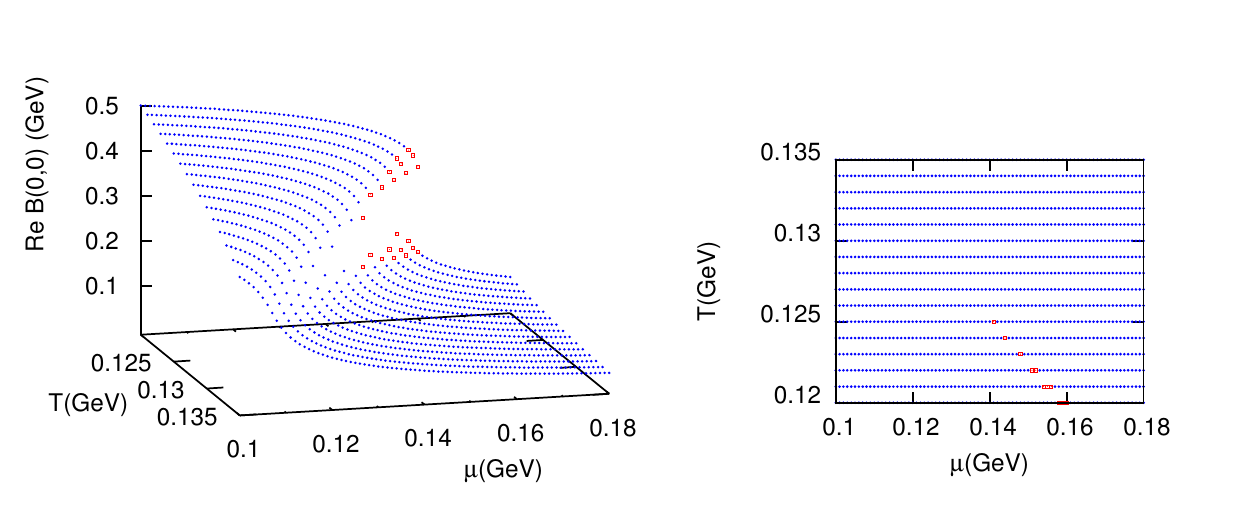}
\caption{\label{fig:2-1} Left: Real part of $B(0,0)$ on the $T\!\!-\!\!\mu$ plane.~~~~~~ Right:
Top view of the left figure.}
\end{figure}

The inclusion of $e^{-\alpha\mu^2/\omega^2}$ does not affect the result at $\mu=0$. The parameter $\alpha$ characterizes the weakening of interaction between quarks with increasing $\mu$. In the perturbative calculations of finite-temperature QCD, the running coupling constant decreases with $\mu$. So we could expect a decrease in the coupling in the nonperturbative region. Thus $\alpha$ controls the rate of decrease. In this paper, we employ this form of suppression to investigate the CEP.

To locate the CEP, we mainly investigate the phase transition near the CEP. This brings convenience in numerical computation: when $T>120 MeV$, we only need to calculate a few Matsubara frequencies. Here, we take 40 Matsubara frequencies into consideration and obtain stable results. The parameters are chosen as $\omega=0.45$ and $\alpha=0.6$, which will be explained later. The solutions are plotted in Fig. \ref{fig:2-1}.

From the behavior of the $B$ function, it can be seen that partial restoration of chiral symmetry  takes place in two ways. At large $T$, when $\mu$ goes up, the Nambu phase solution goes smoothly into Wigner phase, whereas at small $T$, two solutions could coexist at the same $\mu$. We will investigate thermodynamic quantities with these two solutions in the next section.

\section{Critical End Point}
\label{cep}
In this section, we will draw the QCD phase diagram corresponding to Fig.~\ref{fig:2-1}. We use the existence of zero pressure difference between the two phases as the criterion for first order phase transition. CEP is located as the end point of the first order phase transition line. Later, we will show the results of chiral susceptibility as a supplement. Quark condensate is also studied.
\subsection{CJT effective Action and First Order Phase Transition}
\begin{figure}[t]
\begin{minipage}[t]{7cm}
\includegraphics[width=7cm]{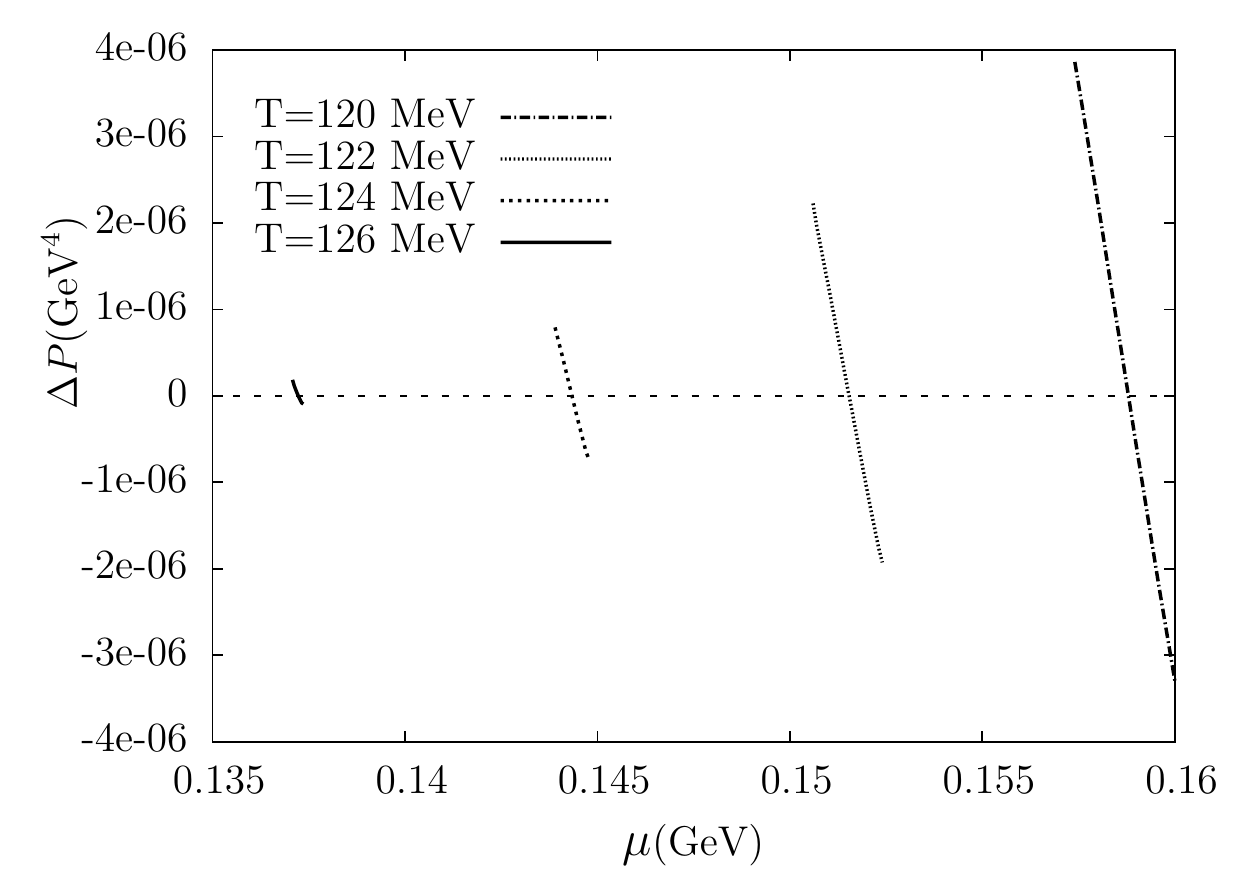}
\caption{\label{fig:3-1-1} Pressure difference between two phases }
\end{minipage}
\hfill
\begin{minipage}[t]{7cm}
\includegraphics[width=7cm]{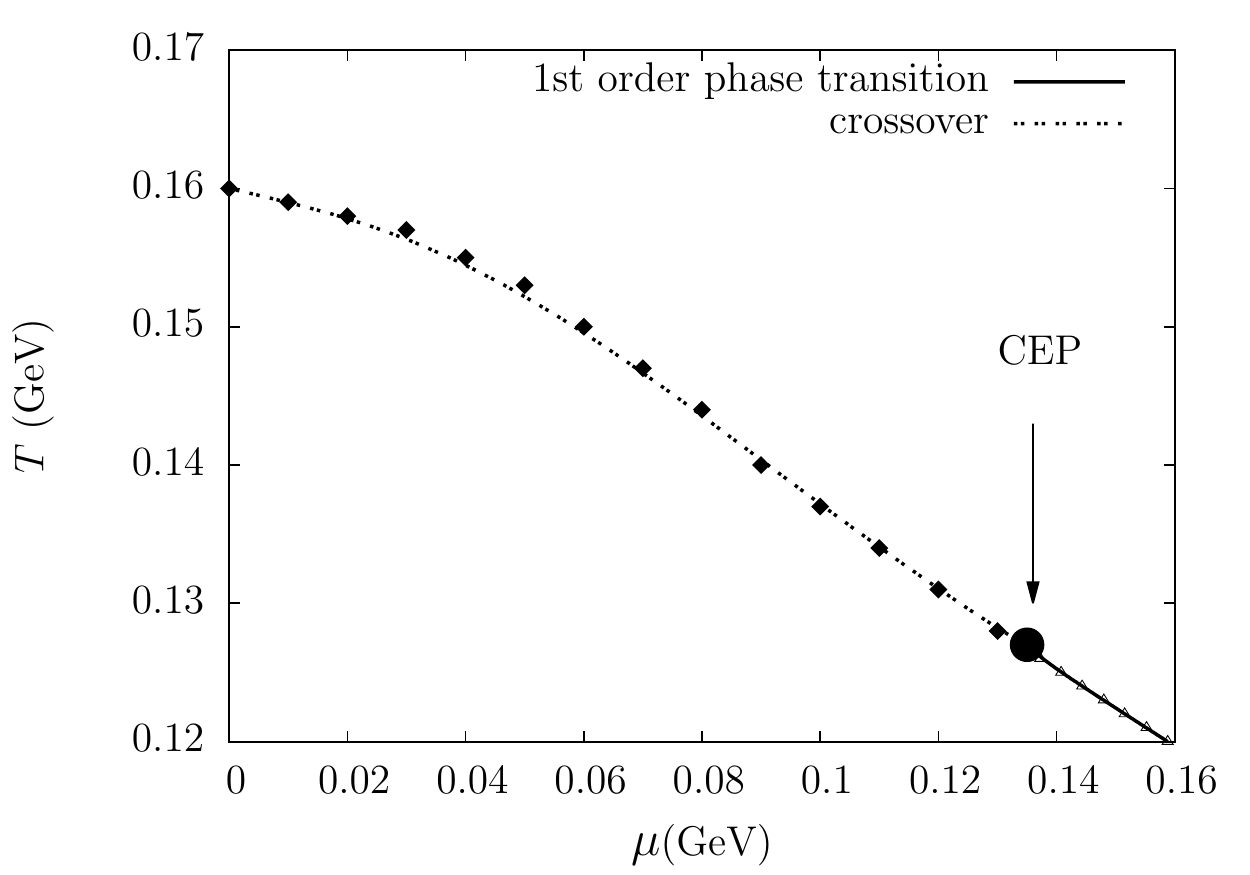}
\caption{\label{fig:3-1-2} QCD phase diagram.}
\end{minipage}
\end{figure}

Cornwall-Jackiw-Tomboulis effective action \cite{n} has been used at finite temperature and density to calculate the partition function \cite{l,x}. Here, we would like to emphasize that it requires that the gluon propagator should not depend on the quark propagator explicitly. Since we are using the rainbow-ladder truncation and a model for the gluon propagator, the CJT effective action is valid here. Actually, in this case the solution of the gap equation is the stationary point for the CJT effective action \cite{z}. So our calculation of the gap equation and the pressure are consistent with each other in this framework. The pressure difference between the two phases is
\begin{equation}
\label{eq:c}
\Delta \mathcal{P}(T,\mu)=\mathcal{P}_N(T,\mu)-\mathcal{P}_W(T,\mu),
\end{equation}
where the subscript $N$ and $W$ stand for $Nambu$ and $Wigner$, respectively. Using the CJT effective action, we have
\begin{eqnarray}
\label{eq:d}
\mathcal{P}_N(T,\mu)&=&\frac{T}{V} Ln \mathcal{Z}(\mu)=\frac{T}{V}\textrm{Tr}\biggr[Ln(G_N^{-1}G_0)-\frac{1}{2}(1-G_0^{-1}G_N)\biggr],
\end{eqnarray}
where the trace operation is over color, flavor, Dirac and coordinate indices all through this paper. By Fourier transformation, we can rewrite Eq. (\ref{eq:d}) in the momentum space along with $\mathcal{P}_W(T,\mu)$ . Then Eq. (\ref{eq:c}) becomes
\begin{eqnarray}
\Delta \mathcal{P}(T,\mu)&=&2 N_c N_f T \sum_{n=-\infty}^\infty \int \frac{d^3 p}{(2\pi)^3}\biggr(Ln\frac{A_N^2 \vec{p}^{~\!2}+C_N^2\tilde{\omega}_n^2+B_N^2}{A_W^2 \vec{p}^{~\!2}+C_W^2\tilde{\omega}_n^2+B_W^2}\nonumber\\
&&+\frac{A_N\vec{p}^{~\!2}+C_N\tilde{\omega}_n^2+B_N m}{A_N^2 \vec{p}^{~\!2}+C_N^2\tilde{\omega}_n^2+B_N^2}-\frac{A_W\vec{p}^{~\!2}+C_W\tilde{\omega}_n^2+B_W m}{A_W^2 \vec{p}^{~\!2}+C_W^2\tilde{\omega}^2_n+B_W^2}\biggr).
\end{eqnarray}
The first order phase transition takes place when $\Delta \mathcal{P}(T,\mu)=0$. The evolution of $\Delta \mathcal{P}(T,\mu)$ with fixed $T$ and increasing $\mu$ is plotted in Fig.~\ref{fig:3-1-1}. Picking out all the points $(T,\mu)$ in the $T-\mu$ plane where $\Delta \mathcal{P}=0$, we can find the first order phase transition line, which is shown in Fig.~\ref{fig:3-1-2}. As we can see, in this case CEP is located at $(T_E,\mu_E)=(0.127\, \textrm{GeV},0.135\, \textrm{GeV})$ .

Here we would like to point out, as can be seen from Fig.~\ref{fig:3-1-1}, while the area for the existence of multi-solution gets narrower with increasing $T$, there is always one point where $\Delta\mathcal{P(\mu)}=0$. The area finally vanishes at the CEP. This fact justifies our use of the criterion for locating CEP due to its consistency with the general phase diagram.

\subsection{Chiral susceptibility and Crossover}
When the chemical potential is lowered to less than $\mu_E$, heating the system would make the Nambu solution change continuously into Wigner solution. In this case, we can use various kinds of susceptibility to characterize what is happening here. One kind of susceptibility, i.e., the chiral susceptibility, is often used in lattice QCD \cite{c}, Nambu-Jona-Lasinio(NJL) model \cite{o,p}, Dyson-Schwinger equation \cite{r,he} and other approaches.

Once the CJT effective action is known, one can in principle derive any thermodynamical quantity from it. In this paper, we will use the CJT effective action to derive directly the chiral condensate and the corresponding chiral susceptibility. The chiral condensate is considered to be the order parameter for the chiral phase transition, which is defined as:
\begin{equation}
\label{eq:cond}
 \langle\bar{\psi} \psi\rangle=-\frac{T}{V} \frac{\partial}{\partial m} Ln \, \mathcal{Z},
\end{equation}
where $\langle\bar{\psi}\psi\rangle=\langle\bar{\psi}\psi\rangle_u+\langle\bar{\psi}\psi\rangle_d$. Substituting Eq.~(\ref{eq:d}) into Eq.~(\ref{eq:cond}), and using $\frac{\partial}{\partial \alpha}\,( \textrm{det} X)= \textrm{det} X \cdot \textrm{Tr}(X^{-1} \frac{\partial X}{\partial \alpha})$, we obtain
\begin{equation}
\label{eq:deriv}
\langle\bar{\psi}\psi\rangle=\frac{T}{V} \textrm{Tr} \biggr[(G-\frac{1}{2} G \, G_0^{-1}\, G)\cdot\frac{\partial}{\partial m} (G^{-1}) +\frac{1}{2} G-G_0\biggr].
\end{equation}
It should be noted that this is a direct definition and works both in the chiral limit and beyond the chiral limit, so we can calculate the quark condensate in our case ($m=5$ MeV) directly. Now it is interesting to compare this definition with the generally used one
\begin{equation}
\label{formercond}
\langle\bar{\psi} \psi\rangle=\frac{T}{V} \textrm{Tr}\, [G-G_0].
\end{equation}
Here, in order to eliminate the ultraviolet divergence, one has subtracted a term $G_0$ by hand (see for example, Ref. \cite{zong}), whereas in Eq. (3.4) the chiral condensate has been obtained directly from the CJT effective action, where this subtraction term appears automatically.

\begin{figure}[t]
\begin{minipage}[t]{7cm}
\includegraphics[width=7cm]{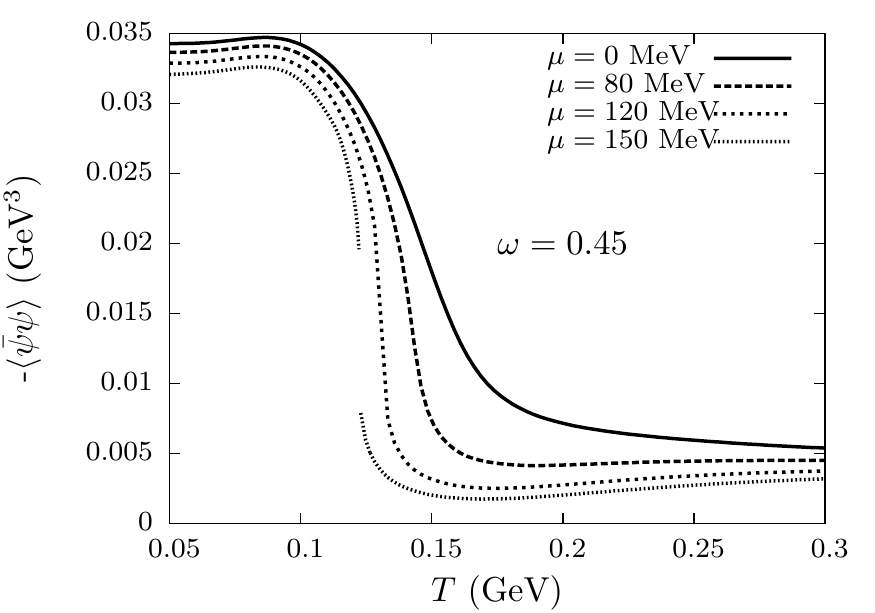}
\caption{\label{fig:3-2-2} Evolution of the chiral condensate}
\end{minipage}
\hfill
\begin{minipage}[t]{7cm}
\includegraphics[width=7cm]{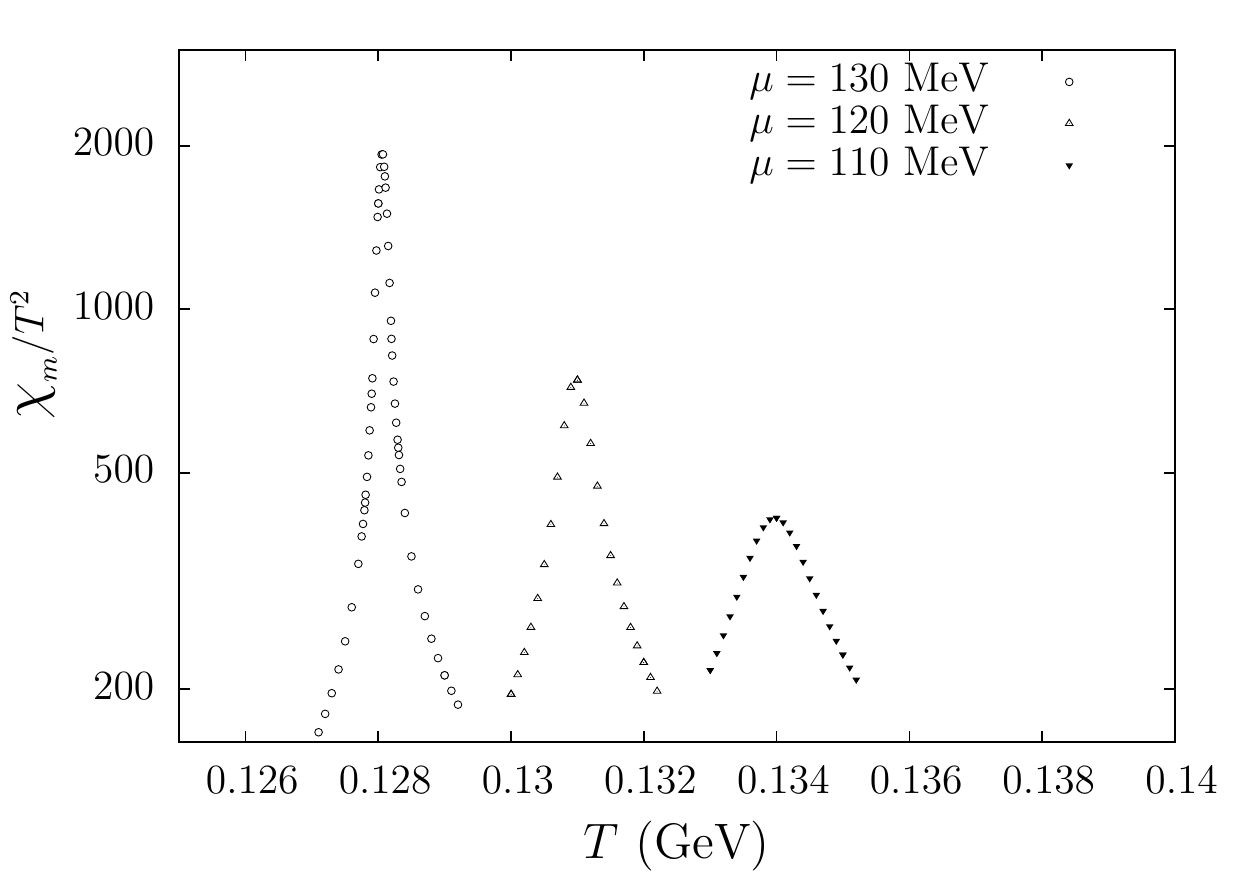}
\caption{\label{fig:3-2-1} Evolution of the chiral susceptibility}
\end{minipage}
\end{figure}

Actually, since we have obtained the effective action in Eq. (\ref{eq:d}), it is easier to take the partial derivative directly. Note that the effective action alone is divergent. What is convergent is the difference of two properly chosen effective actions. For our purpose here, we choose two effective actions with two very close current quark masses $m$ and $m^\prime$. The calculation of Eq.~(\ref{eq:cond}) is then straightforward. The result is shown in Fig.~\ref{fig:3-2-1}. As we can see, when $\mu<\mu_E$ the quark condensate decreases continuously as $T$ goes up. This can be compared with the case  $\mu>\mu_E$ (e.g., 150 MeV) where the quark condensate drops discontinuously at some point. So, a first order phase transition can be expected here.

The chiral susceptibility can be treated similarly. It is defined as
\begin{equation}
\label{eq:sus}
\textrm{{\Large $\chi$}}_m=\frac{T}{V}\frac{\partial^2}{\partial m^2}Ln \mathcal{Z}=-\frac{\partial}{\partial m}\langle\bar{\psi}\psi\rangle
\end{equation}
Taking the second order partial derivative of the CJT effective action, we can obtain Fig.~\ref{fig:3-2-1}. It can be seen that when $\mu < \mu_E$, the chiral susceptibility undergoes a smooth change with increasing $T$. This is where crossover takes place. When approaching the CEP, the curve gets sharper. The peak tends to go to infinity near the CEP, which is a second order phase transition point. We pick out the coordinates of the maximums and obtain the dashed line in Fig.~\ref{fig:3-1-2}.

\subsection{Conclusions}
\begin{figure}[!h]
\begin{minipage}[t]{7cm}
\includegraphics[width=7.0cm]{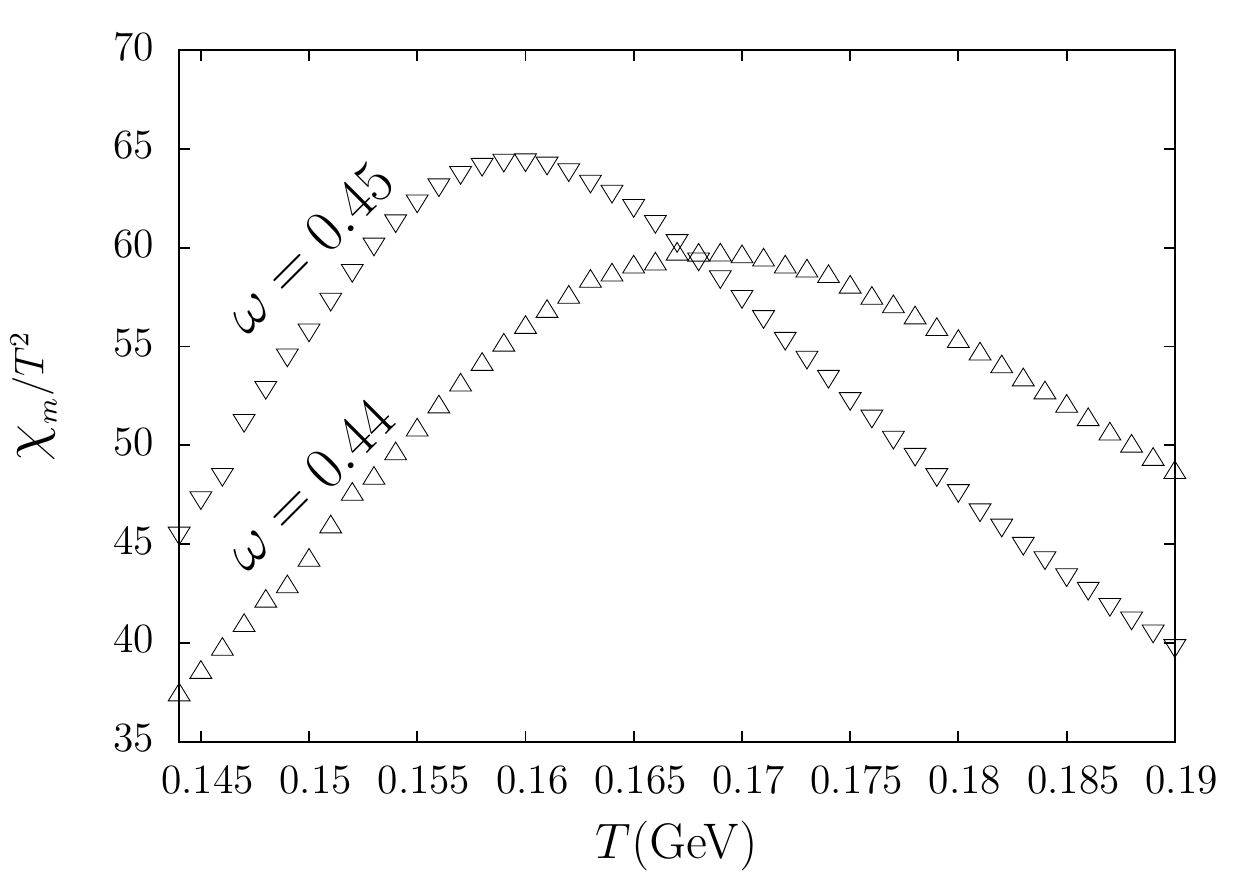}
\caption{\label{fig:3-3-1}Chiral susceptibility evaluated with different $\omega$ at $\mu=0$}
\end{minipage}
\hfill
\begin{minipage}[t]{7cm}
\includegraphics[width=7.0cm]{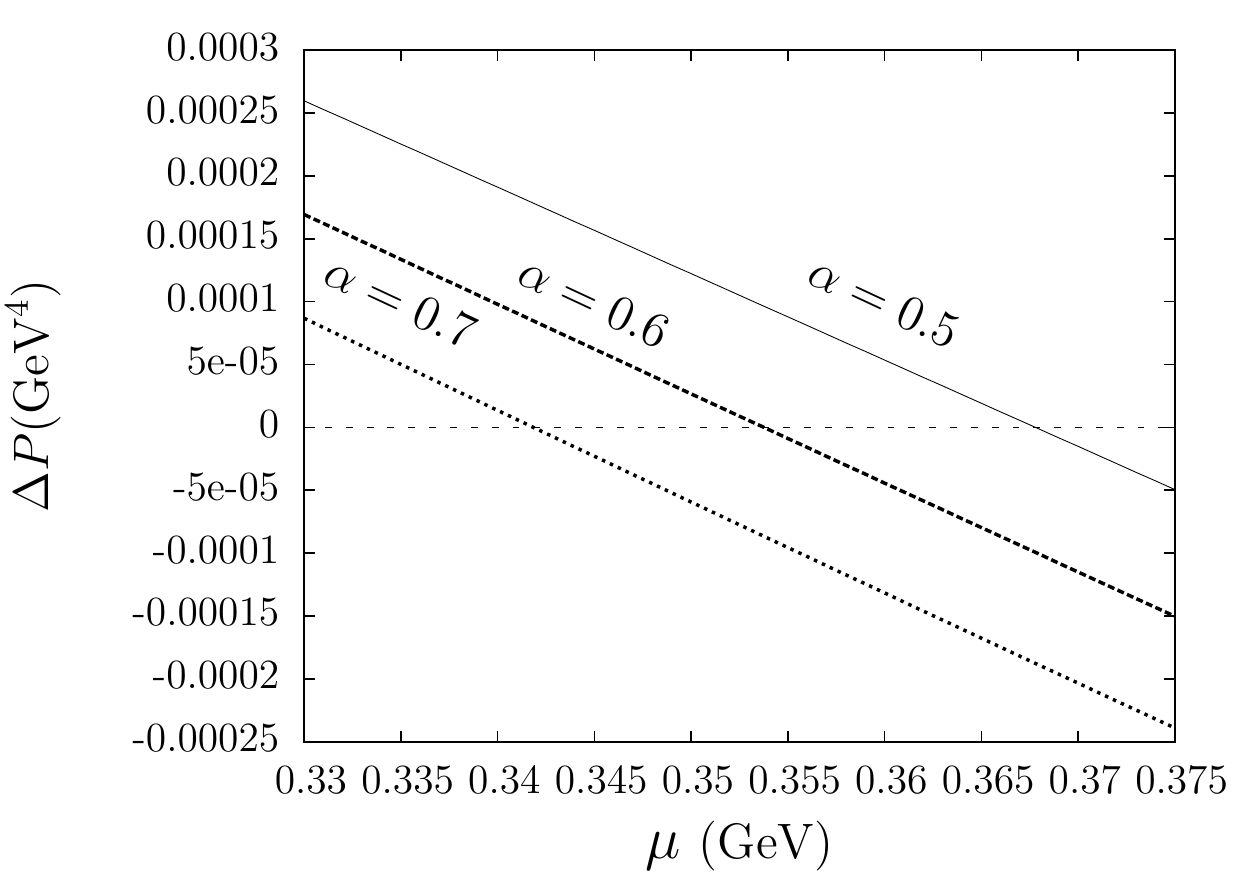}
\caption{\label{fig:3-3-2} Pressure difference evaluated with different $\alpha$ at $T=0$}
\end{minipage}
\end{figure}

Taking $m=5$ MeV, we are left with two independent parameters: $\omega$ from the Maris-Tandy model and $\alpha$ which controls the decrease of coupling with respect to increase of $\mu$. Note that $\alpha$ does not  affect the results when $\mu=0$. Lattice QCD has given many results in this case \cite{t,u,f,ac}. To locate the CEP in our work, we adopt the recent lattice result $T_c \simeq$157 MeV \cite{ac}. As can be seen from Fig.~\ref{fig:3-3-1}, in our case $\omega=0.45$ reproduces $T_c\simeq160$ MeV.  Here we need to point out that we should be careful when fixing the  $\omega$. Looking back at Eq.~(\ref{eq:MT}), one should notice there is a factor $D_0/\omega$ describing the magnitude of the coupling strength. So there is actually a seventh power of $\omega$ in the denominator. When treating hadron physics, which corresponds to zero temperature, D$\chi$SB is easily achieved and research has focused on hadron properties.  However, when it comes to QCD phase diagram study, the strength of coupling becomes important, because we know only a strong enough coupling could cause D$\chi$SB. So the value of $\omega$ should guarantee that $T_c$ is centered in recent lattice results. After fixing $\omega$, we try to find the appropriate value for $\alpha$. At zero temperature, when $\mu$ goes up to 1/3 of baryon mass, nuclear matter turns up and quark number density becomes nonzero\cite{u1}. It is generally believed that the chiral phase transition point must be larger than 1/3 of baryon mass. Here, we refer to the $\mu_c$ at zero temperature given by various studies \cite{a,aa,ab} and spread the results over a possible region, in our case, from about 0.340 GeV to 0.370 GeV. The values of $\alpha$ and the relevant $\mu_c$ can be seen from Fig.~\ref{fig:3-3-2}.

Table.~\ref{tab:3-3-3} shows our results for different values of $\alpha$. At first sight it is surprising that $\mu_E$ goes up as $\alpha$ increases. This is quite different from the case of $\mu_c$. However, if one takes a closer look, one will find a decrease in $T_E$, which actually compensates for the increase of $\mu_E$. Hence, the total effect is that increasing $\alpha$ would make the CEP rotate clockwise. In general, our result shows CEP is located at the lower boundary of the crossover band given by Ref. \cite{ac}. This is like other model studies that give a relatively low $T_E/T_c$ value \cite{w}. Note that the fourth column in Table.~\ref{tab:3-3-3} gives a $(\mu_E,T_E)/T_c$ around (0.78,0.88). it is close to that of Polyakov-quark-meson model which has a $(\mu_E,T_E)/T_c$ around (0.81,0.91) \cite{ad}.  Considering we only used Maris-Tandy model here, it could be interesting to see what different gluon propagator models would exhibit within Dyson-Shwinger equations approach.

\begin{table}[!h]
\centering
\begin{tabular}{|c|c|c|c|}
\hline
$\alpha$& 0.5&0.6&0.7\\
\hline
$\mu_c$ /GeV &0.367&0.354&0.342\\
\hline
$(T_E,\mu_E)$ /GeV&(0.129,0.130)& (0.127,0.135) &(0.124,0.141) \\
\hline
\end{tabular}
\caption{\label{tab:3-3-3} Results for different $\alpha$}
\end{table}

\section{Summary}
In this paper we generalize the study in Ref.  \cite{i} to investigate the QCD phase diagram at finite temperature and density. The gap equation is solved beyond chiral limit and an area for the existence of multi-solution is found. We then use CJT effective action which is consistent with our gap equation to calculate all the thermodynamical quantities. Although the CJT effective action alone is divergent, we could obtain convergent results for pressure difference, quark condensate and chiral susceptibility. The first order phase transition and crossover are both studied, and from this the CEP is identified and located.

\acknowledgments
We thank P. C. Tandy and C. D. Roberts for helpful discussions. This work is supported in part by the National Natural Science Foundation of China (under Grant 11275097, 10935001, 11047020, 11274166 and 11075075), the National Basic Research Program of China (under Grant 2012CB921504) and the Research Fund for the Doctoral Program of Higher Education (under Grant No 2012009111002).

\end{document}